\documentclass[aps,preprint,onecolumn,showpacs,superscriptaddress,groupedaddress,nofootinbib]{revtex4}  
\usepackage{graphicx}  
\usepackage{dcolumn}   
\usepackage{bm}        
\usepackage{amssymb}   
\usepackage{color}

\renewcommand{\vec}{\boldsymbol}
\newcommand{\be}{\begin{equation}}
\newcommand{\ee}{\end{equation}}
\newcommand{\bear}{\begin{eqnarray}}
\newcommand{\eear}{\end{eqnarray}}
\newcommand{\ba}{\begin{array}}
\newcommand{\ea}{\end{array}}
\newcommand{\ctwo}{\Delta v_2(0)}
\newcommand{\taufree}{\tau_{\rm free}}

\begin{document}

\title{Charged elliptic flow at zero charge asymmetry}

\author{Misha Stephanov}
\email{misha@uic.edu}
\affiliation{Department of Physics, University of Illinois, Chicago, Illinois 60607, USA}

\author{Ho-Ung Yee}
\email{hyee@uic.edu}
\affiliation{Department of Physics, University of Illinois, Chicago, Illinois 60607, USA}
\affiliation{RIKEN-BNL Research Center, Brookhaven National Laboratory, Upton, New York 11973-5000 }

\date{\today}

\begin{abstract}

  The difference between the flow ellipticities of oppositely charged
  pions, $\Delta v_2\equiv v_2[\pi^-]-v_2[\pi^+]$, measured recently
  by STAR collaboration at RHIC shows a linear dependence
  on the event charge asymmetry $A_\pm\equiv {(N_+-N_-)/ (N_++N_-)}$:
  $\Delta v_2(A_\pm)= \ctwo+ r A_\pm$ with a slope $r>0$ and a
  non-zero intercept $\ctwo>0$ of order $10^{-4}$. We
  point out two novel mechanisms which could explain the non-zero
  value of the charged elliptic flow $\Delta v_2$ at {\em zero}
  charge asymmetry $A_\pm=0$, i.e., the non-zero positive intercept
  $\ctwo$. Both effects are due to the electric fields created by the
  colliding ions. These fields have quadrupole asymmetry of the magnitude
  and the sign needed to account for the nonzero intercept $\ctwo>0$
  in the RHIC data. One of the mechanisms also involves chiral magnetic
  effect. This mechanism, although negligible at RHIC energies, may become
  important at LHC energies.

\end{abstract}
\preprint{RBRC 975}
\maketitle

\section{Introduction}

The recent analysis~\cite{Wang:2012qs,Ke:2012qb} of STAR
event-by-event measurements of the difference between the elliptic
flows of positive and negative pions, $\Delta v_2\equiv
v_2[\pi^-]-v_2[\pi^+]$, finds an interesting behavior of this quantity
as a function of the event charge asymmetry $A_\pm\equiv
{(N_+-N_-)/ (N_++N_-)}$.  As seen in FIG. \ref{fig1}, $\Delta v_2$
depends linearly on $A_\pm$, i.e.,
\begin{equation}
\label{eq:delta-v2}
\Delta v_2(A_\pm)=\ctwo+r A_\pm\,,
\ee
with a slope $r>0$ and an intercept $\ctwo>0$. 
\begin{figure}[h]
	\includegraphics[width=8cm]{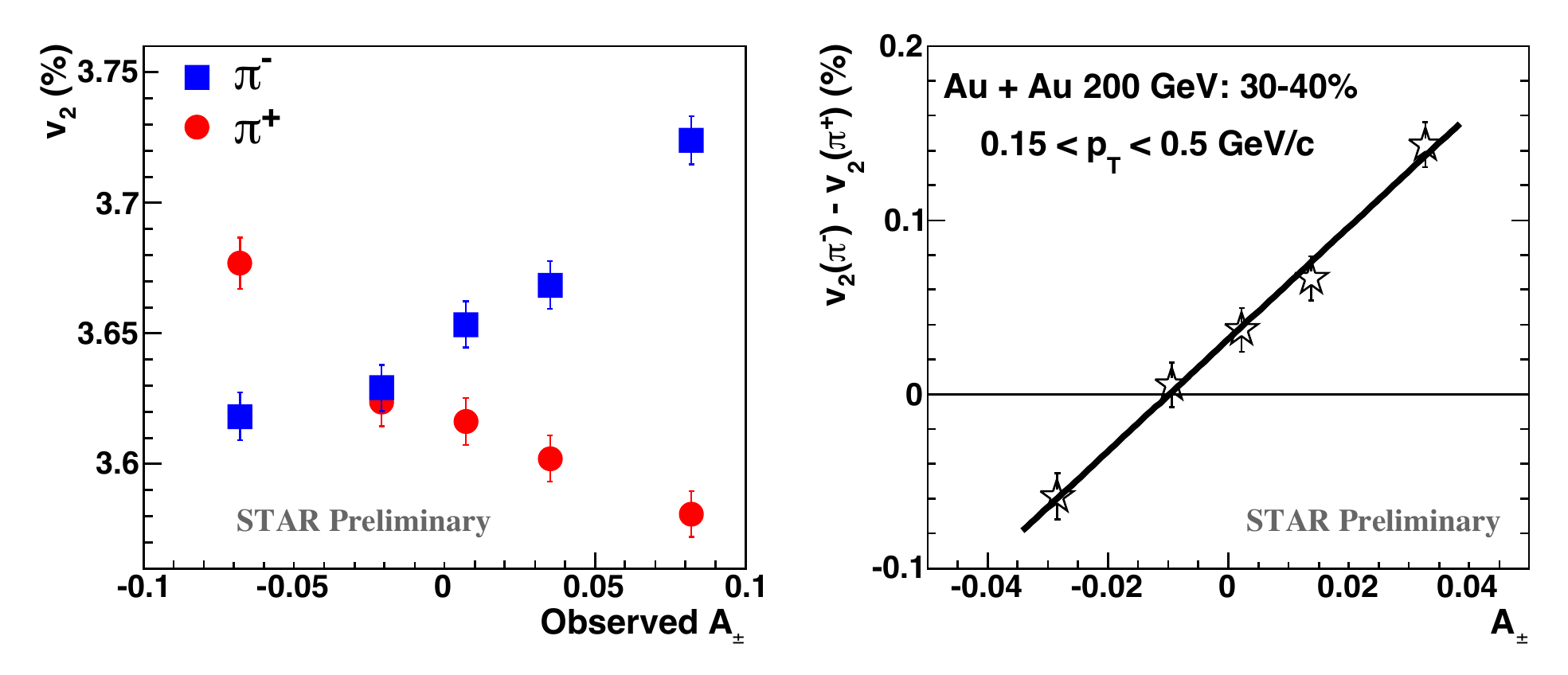}
		\caption{\label{fig1}The STAR data \cite{Wang:2012qs} on the elliptic flow difference between the positive and negative pions showing a linear dependence on the charge asymmetry $A_\pm\equiv 
		{(N_+-N_-)/ (N_++N_-)}$ with a finite positive intercept.}
\end{figure}
This observed positive slope agrees with the prediction made in Ref.~\cite{Burnier:2011bf,Burnier:2012ae} based on the Chiral Magnetic Wave (CMW)~\cite{Kharzeev:2010gd,Newman:2005hd}.
The CMW is a gapless sound-like mode of chiral charges (left-handed or right-handed) propagating along the direction of the magnetic field, with a dispersion relation
\begin{equation}
\omega=\pm v_\chi k-iD_L k^2+\cdots\,,\label{dispersion}
\ee
where the sign, and hence the direction of the propagation with respect to the magnetic field, depends on the chirality of the charge fluctuations.
The velocity $v_\chi$ is given by \cite{Kharzeev:2010gd}
\begin{equation}
v_\chi={N_c e B\over 4\pi^2}\alpha\,,
\ee 
where  $\alpha$ is the  inverse susceptibility of the chiral charge,
and  $D_L$ -- the longitudinal diffusion constant which depends on the
microscopic transport properties of a given system. 
The mechanism described in Ref.\cite{Burnier:2011bf,Burnier:2012ae} is
based on the
observation (also made in Ref.\cite{Gorbar:2011ya}) that the initial {\em net} electric charge $Q$ is a sum of the
net charge carried by left-handed ($Q_L$) and right-handed ($Q_R$) chiral
carriers (quarks):
\begin{equation}
Q=Q_L+Q_R\,,
\ee 
where on average $Q_R=Q_L$.
In off-central heavy-ion collisions at RHIC energies the heavy ions can create a transient magnetic field of strength as large as $B\sim 4 m_\pi^2/e \sim 10^{19}{\rm G}$ in the direction perpendicular to the reaction plane.
The CMW moves chiral charges $Q_L$ and $Q_R$ along the axis of this magnetic field according to Eq.~(\ref{dispersion}). Since $Q_L$ moves
in the direction opposite to that of $Q_R$, 
the net result will be 
an electric charge excess in the poles of the fireball and a depletion in the central region. 
\begin{figure}[h]
	\includegraphics[width=5cm]{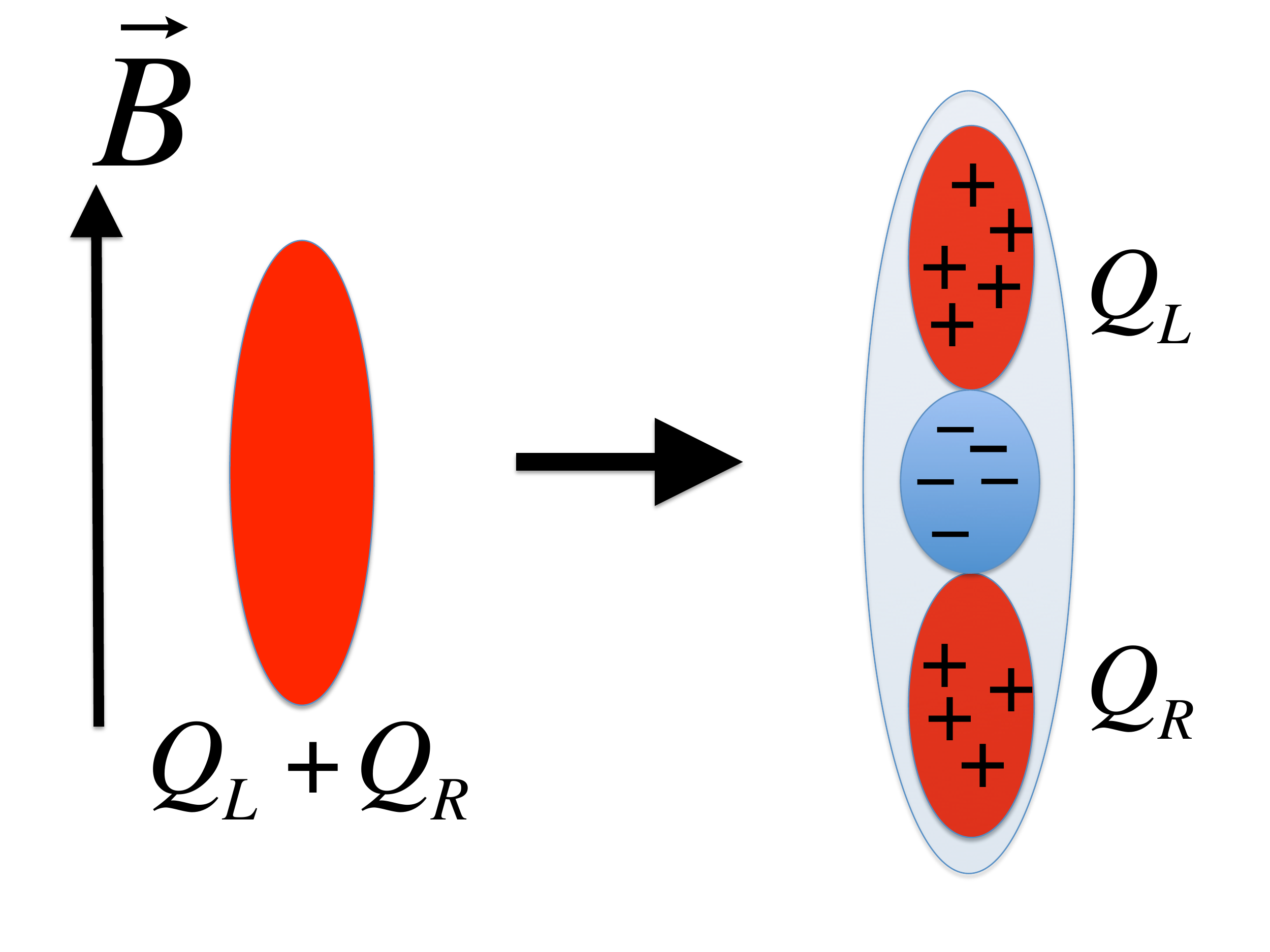}
			\caption{\label{fig2} The Chiral Magnetic Wave
                          (CMW) evolution of the initial charge asymmetry
                          (taken to be positive in this example)
                          produces a net electric quadrupole moment
                          responsible for a non-zero $\Delta v_2\equiv
                          v_2[\pi^-]-v_2[\pi^+]$ (positive in this
                          this case). The collision plane
                          is horizontal in this sketch.}
\end{figure}
See FIG. \ref{fig2}
for a schematic illustration of the mechanism.  
Such a charge transport via CMW will result in an electric quadrupole moment whose magnitude is naturally
proportional to the charge asymmetry. Combined with the subsequent
radial flow this leads to
the charged elliptic flow of pions, $\Delta v_2$, linear in $A_\pm$ \cite{Burnier:2011bf,Burnier:2012ae}.  
A semi-realistic theoretical simulation performed in
Ref.\cite{Burnier:2011bf,Burnier:2012ae} qualitatively and semiquantitatively agrees with
the STAR data, including a non-trivial dependence on the impact parameter (centrality).

Although the CMW mechanism described above accounts for the
nonzero positive slope in Fig.~\ref{fig1}, the origin of non-zero
positive intercept $\ctwo$ has not been so far understood to the same extent. {An argument based on a simple quark coalescence model in
  Ref.\cite{Dunlop:2011cf} suggests that
  a positive contribution to the intercept may appear due to initial
  isospin asymmetry.}
In this paper we shall identify two different mechanisms which are more
closely related to the CMW effect responsible for the nonzero slope.
These mechanisms account for
the sign as the well as the magnitude of the intercept.

One way to look at the problem is to note that the nonzero intercept
suggests the existence of an additional source of
electric quadrupole moment, approximately independent of the charge
asymmetry $A_\pm$ and present even in the
neutral plasma, i.e., at $A_\pm=0$.
We point out two sources of such a net positive
(out of reaction plane) electric
quadrupole moment. 
Both involve electric fields created by the
heavy ions which are similar in strength to the magnetic fields
responsible for the CMW. We describe the two mechanisms in the
subsequent sections and make an order of magnitude estimate of their effect on
the charged elliptic flow  $\Delta v_2(0)$.

\section{Source 1: Electric field quadrupole}

A highly relativistic heavy ion carries around itself a Lorentz contracted
``pancake'' of the electromagnetic field perpendicular to its velocity.
The magnetic field lines are circular around the velocity direction
and the heavy ions colliding at
 a finite impact parameter create a net magnetic field in the
 overlapping region along the direction perpendicular to the reaction
 plane. This field is responsible for the CMW. 
\begin{figure}[h]
	\includegraphics[width=4cm]{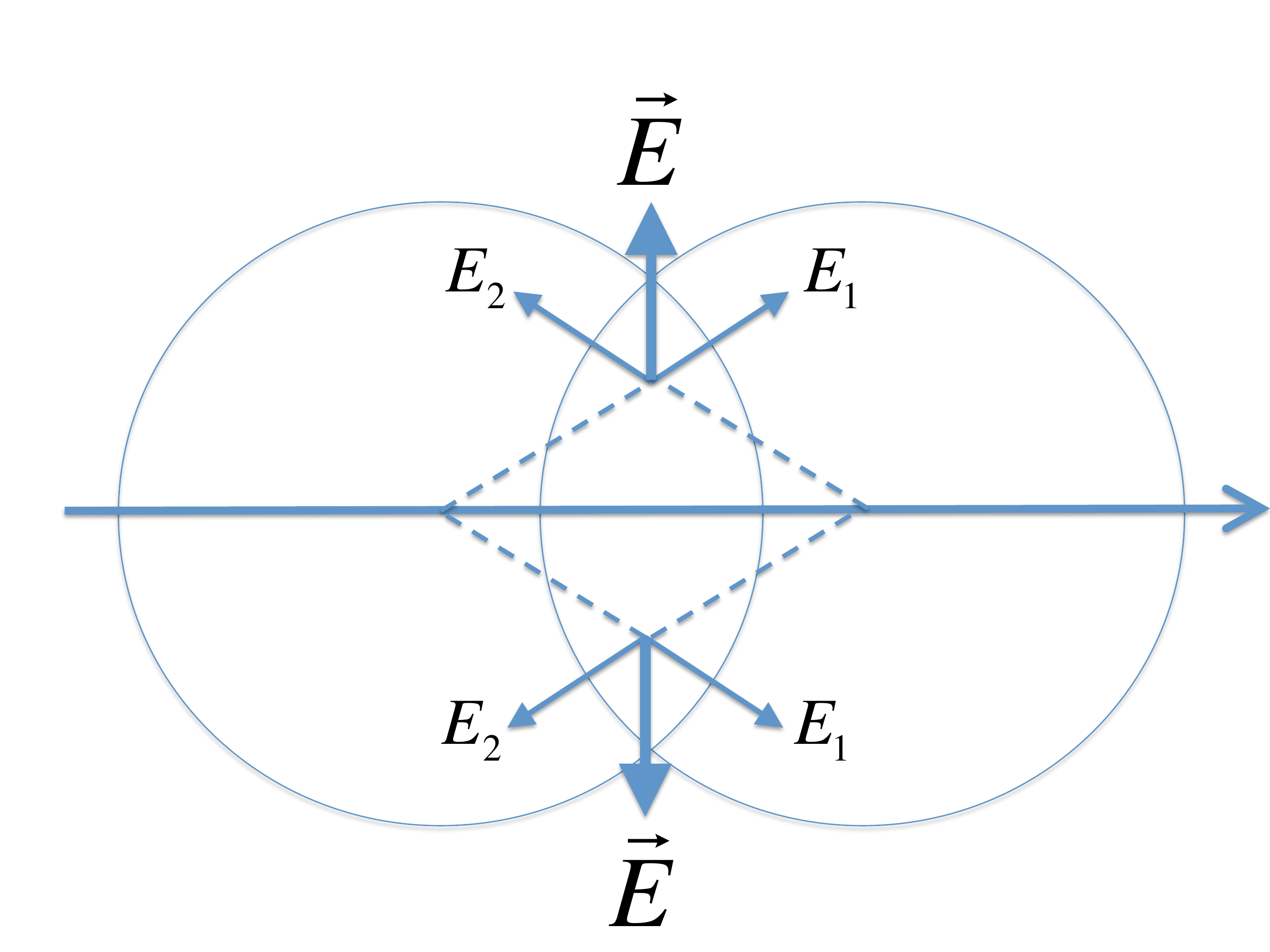}\includegraphics[width=4cm]{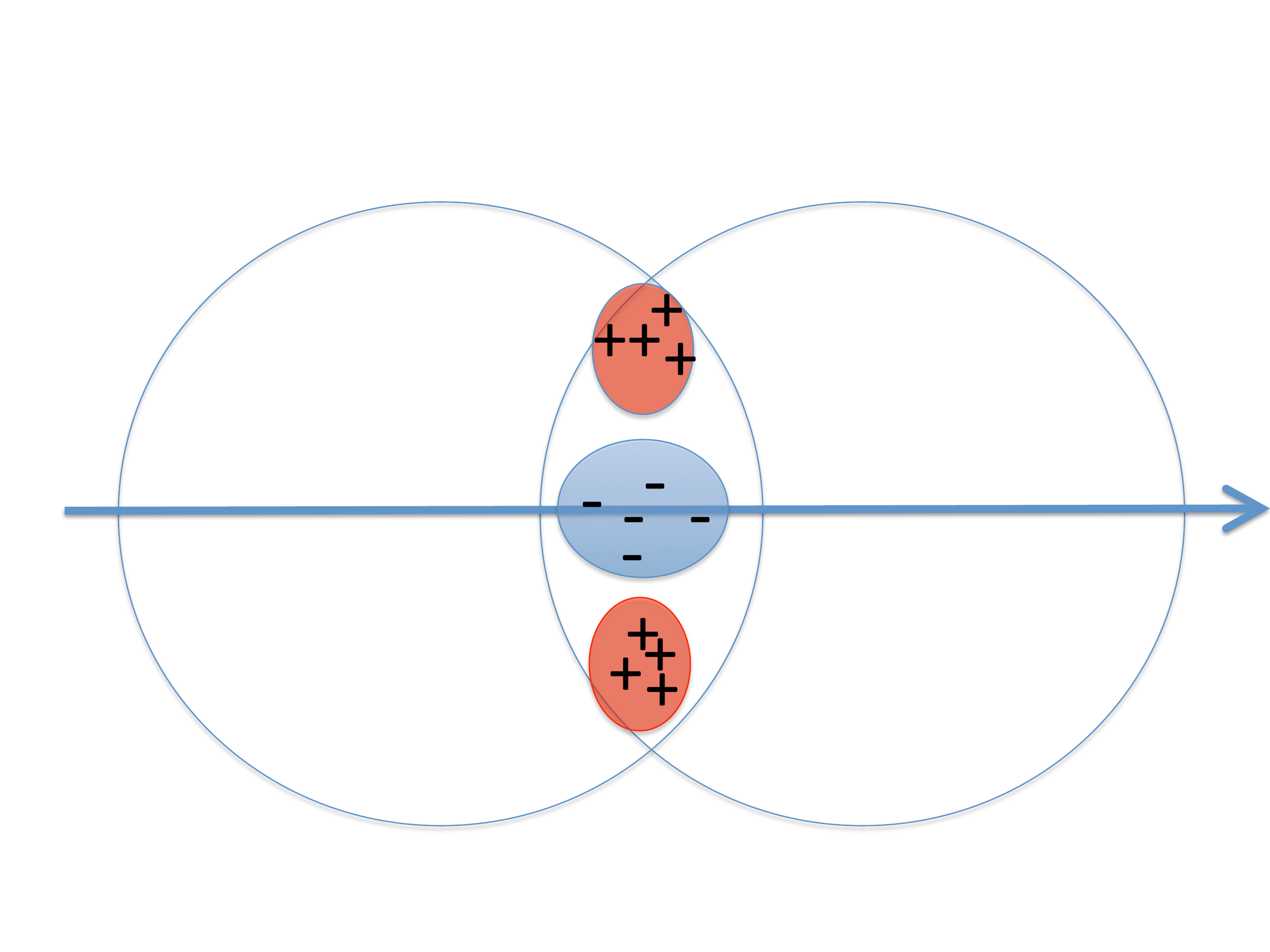}
		\caption{\label{fig3}The net electric field profile from the two heavy-ions ({\it left}), which leads to a non-zero electric quadrupole moment ({\it right}) whose magnitude
		is independent of the charge asymmetry $A_\pm$ for
                small enough $A_\pm$. The reaction plane is horizontal.}
\end{figure}

On the other hand, the electric field from a single heavy ion is radially directed, perpendicular to the magnetic field at all points in space-time.
The superposition of the electric fields from the two colliding ions
in the overlapping region is described schematically in FIG.~\ref{fig3}(left).
The key feature of the effect we describe is that this electric field
has a quadrupole component: the field
pointing outward {\it perpendicular to} the reaction plane is stronger than the field
pointing outward {\it in} the reaction plane due to partial cancellation of
the in-plane components.
Since the plasma (or pre-equilibrium state, such as glasma, at such
early times) is conducting, there naturally arise net currents
moving charges outward from the reaction plane: $\vec J \sim \vec E$. 
One easily sees that this charge transport induces a net positive
electric quadrupole moment along the out-of-plane axis (see
FIG. \ref{fig3}) which, combined with subsequent radial
collective flow, gives rise to $\Delta v_2>0$ as in the CMW.

Since the conductivity is non-zero also in a {\em neutral} plasma, the effect is
present even when $A_\pm=0$. It is also clear from charge conjugation
symmetry that the conductivity can only depend quadratically on the
net charge density. Thus the dependence  of the contribution of the
quadrupole electric field effect to $\Delta v_2$ in
Eq.~(\ref{eq:delta-v2}) on $A_\pm$ is given by $\ctwo+{\cal
  O}(A_\pm^2)$. The term quadratic in $A_\pm$ is negligible for small  enough $A_\pm$ ($A_\pm$ in FIG. \ref{fig1} is $\le 0.04$). 

We emphasize again that the magnitude of the electric field is similar to that of the magnetic field, which indicates that 
the order of magnitude of the induced quadrupole moment should be comparable to the one induced by the magnetic field via the CMW.
This aspect and other similarities make this mechanism (and also the mechanism described in the following section) to be a natural possible 
explanation of the non-zero intercept in FIG.~\ref{fig1}, if the nonzero slope is due to the CMW.

Let us make an order of magnitude estimate of the effect. 
The elliptic flow difference between positively and negatively charged
particles is of order $\Delta v_2\sim {\Delta Q/ N_{\rm tot}}$ where $\Delta Q$  is
the total charge  (in units of $e$) transported by this mechanism and $N_{\rm tot}$ is the total multiplicity of charged particles.
The value of $\Delta Q$ can be estimated as $\Delta Q\approx J A
\tau_J$, where $J$ is the (units of $e$) 
current density induced by the electric
field, $\tau_J$ is the lifetime of the current and $A$ is the typical area of
the fireball transverse to current during the lifetime of the electric
field, $\tau_E$. The area transverse to the current at time $\tau_E$ is
given by the product of the transverse size, which for $30-40$\%
central events we take to be of
order of the radius of the nucleus $R\sim 7$ fm, and of the longitudinal size, which
is of order $\tau_E$: $A=R\tau_E$.

To estimate the current and its lifetime we need to consider two
cases depending on whether the lifetime $\tau_E$ of the electric field
is shorter or longer than the typical  (mean
free) time between collisions
$\taufree$. Although the mechanisms of the charge transfer are
different in the two cases, the relevant product, $J\ \tau_J$, turns
out to be similar. As we shall see below, for relatively
long-lived electric field, $\tau_E\gg\taufree$, the magnitude of the current is
determined by conductivity and
$J\sim \taufree$, while its lifetime is as long as that of the
electric field $\tau_J\approx\tau_E$. On the other hand,
for short-lived electric field, $\tau_E\ll\taufree$, the magnitude of
the current is
determined by the lifetime of the field
$J\sim \tau_E$, while its lifetime is limited by the mean free time  $\tau_J\approx\taufree$. In
either case the relevant product  $J\tau_J\sim \tau_E\taufree$.

Let us now estimate the current $J$. First consider the case when
the lifetime of the electric field $\tau_E$ 
is much longer than the time between collisions
$\taufree$. This is the
hydrodynamic regime and thus the current is given by Ohm's law $J=\sigma E$, where 
for the conductivity $\sigma$ we could use a lattice QCD result from Ref.~\cite{Ding:2010ga},
\begin{equation}
  \sigma=0.4 e T\sum_{F=u,d,s} q_F^2\approx 0.27 eT\,.\label{conductivity}
\end{equation}

  When $\tau_E\ll \taufree$
  we cannot use Ohm's law. But in this case we could estimate the
  current in the collisionless approximation. The
contribution of a each quark or antiquark species $f$ to  the
current is given by $J_f=q_f n_f v_f$ (where $f=u,d,s,\bar u,\bar d,\bar s$ ) where $v_f$ is the drift velocity induced by
the electric field by increasing momentum density of this
component of the plasma by $\pi_f=\tau_E\,eE q_f n_f$. By Lorentz invariance $v_f=\pi_f/w_f$,
where $w_f$ is
the momentum "susceptibility" (given by enthalpy $\epsilon+p$ in
equilibrium). Putting this together, and taking into account that, at $A_\pm=0$,
quark and antiquark of a given flavor contribute equally to the net
current, i.e., $J_F=J_f+J_{\bar f}=2J_f$, we find: $J = \sum_{F=u,d,s} J_F = 
2 eE\tau_E \sum_{F=u,d,s} q_F^2 n_F^2/w_F$, where the factor 2 in the
last equation counts the equal contributions from quarks and
antiquarks of a given flavor $F$.
We can estimate $n_F$ and $w_F$
using equilibrium momentum distribution for massless quarks:
$n_F=9\zeta(3) T^3/\pi^2$ and $w_F=7\pi^2T^4/15$.
Thus we find
\begin{equation}
  \label{eq:jE}
  J\approx  0.26\, \tau_E  eET^2 \sum_{F=u,d,s} q_F^2\,.
\end{equation}
Comparing to Eq.~(\ref{conductivity}) we can write $J\approx
0.65(\tau_E T)\sigma E$.
Clearly, when $\tau_E\approx\taufree$, the two equations should
match, from which we could infer a rough estimate of the mean free
 time $\taufree\approx 1.5/T$ implicit in the lattice result
Eq.~(\ref{conductivity}). Using this as an estimate of the $\taufree$, we can write the result in a form independent of which regime
we consider:
\begin{equation}
  \label{eq:jtau}
  J\tau_J \approx
0.17 eET^2\tau_E\taufree\,.
\end{equation}
Thus
\begin{equation}
  \label{eq:dv2}
  \Delta v_2 \approx \frac{\Delta Q}{N_{\rm tot}} \approx 
\frac{J\tau_J A}{N_{\rm tot}}\approx 0.17 \,\frac{eE\,T^2\tau_E^2\taufree R}{N_{\rm tot}}\,.
\end{equation}

The electric field strength can be approximated as $eE\approx {\gamma\,Ze^2/( 4\pi R^2)}$ where $Z$ is the atomic number of the heavy nucleus, $R$ is its radius, and $\gamma={\sqrt{s}/( 2\,\, {\rm GeV})}$
is the Lorentz contraction factor which we express in terms of the
center of mass energy per colliding nucleon pair. The life-time of the electric field
is roughly given by the longitudinally contracted thickness of the
colliding nucleus, which is $\tau_E\approx {2R/\gamma}$, while 
$\taufree\approx 1.5/T$ as inferred from the lattice result
(\ref{conductivity}). Putting this together and using $Z\approx 80$
and $R\approx 7\, {\rm fm}$ we find
\begin{equation}
  \label{eq:dv2-2}
  \Delta v_2 \approx 1.\times \frac{Z\alpha RT}{N_{\rm tot}\gamma}
\approx 10^{-4} 
\left(T\over 400\, {\rm MeV}\right)
\left(10^3\over N_{\rm tot}\right)
\left(200\,{\rm GeV}\over\sqrt{s}\right)\,,
\end{equation}
which has the right order of magnitude to account for the intercept
$\ctwo\approx 3.\,10^{-4}$ in Fig.~\ref{fig1}.

\section{Source 2: Electric quadrupole via chiral magnetic effect }

The other source of a net positive electric quadrupole moment arises through a combination of both $\vec E$ and $\vec B$.
Consider a P- and CP-odd quantity $\vec E\cdot\vec B$. From the electric field pattern in FIG. \ref{fig3} as well as the magnetic field pointing upward perpendicular to the reaction plane,
one easily recognizes that $\vec E\cdot\vec B$ is positive in the
upper half region, while it is negative in the lower half region as shown in FIG. \ref{fig4}(left).
Via the triangle anomaly relation for the axial current $J_A^\mu$,
\begin{equation}
\partial_\mu J_A^\mu={N_c e^2\over 2\pi^2}\left(\sum_F q_F^2\right)\vec E\cdot\vec B\,,
\end{equation}
this implies a net positive axial charge created in the upper region,
and net negative in the lower region.
In the presence of the vertical magnetic field, the Chiral Magnetic Effect  \cite{Kharzeev:2007jp,Fukushima:2008xe,Son:2004tq} acting on these axial charges induces an electric charge current (in units of $e$),
\begin{equation}
\vec J={N_c e\over 2\pi^2}\left(\sum_F q_F^2\right)\mu_A \vec B\,,\label{CME}
\end{equation}
from which it is easily seen that the net effect is a development of a
positive electric quadrupole moment. See FIG. \ref{fig4} (right) for a schematic explanation.
To some extent, the mechanism is similar to that of the CMW in
Ref.\cite{Burnier:2011bf,Burnier:2012ae}, with a main difference being
that the mechanism is independent of
the initial charge asymmetry $A_\pm$.
\begin{figure}[h]
	\includegraphics[width=4cm]{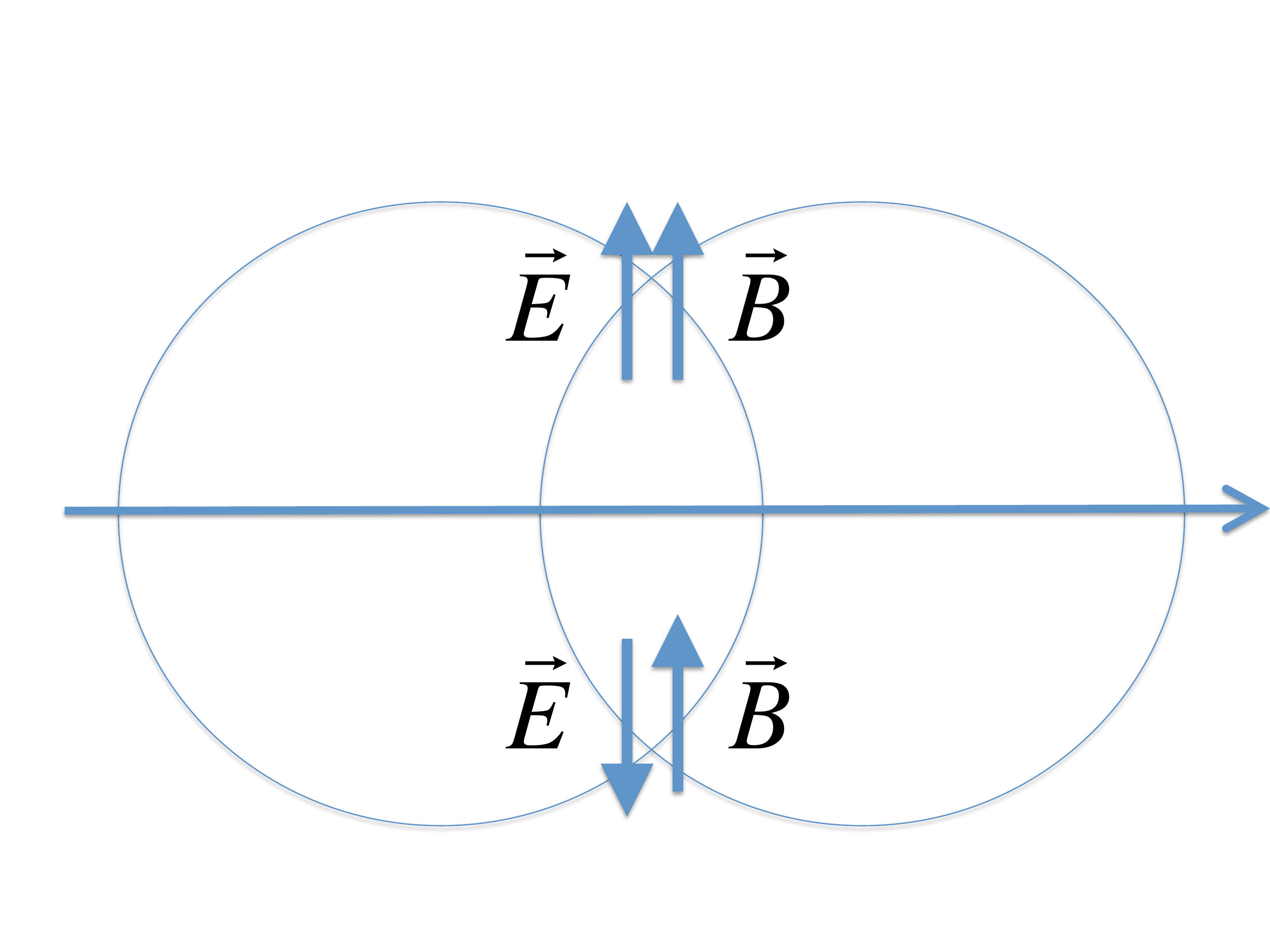}\includegraphics[width=4cm]{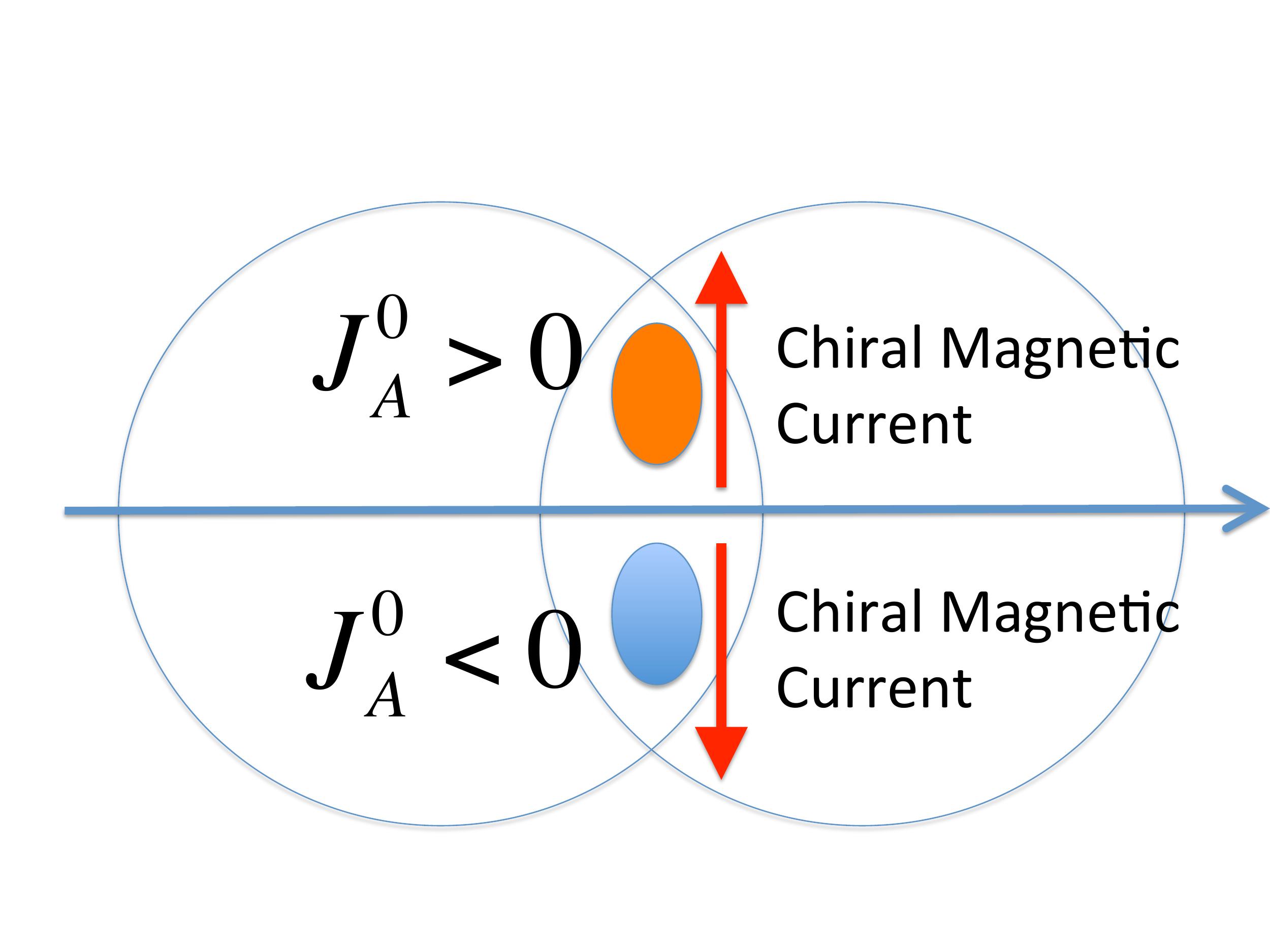}
		\caption{\label{fig4}The profile of $\vec E$ and $\vec B$ that leads to a creation of axial charges via triangle anomaly ({\it left}). The Chiral Magnetic Effect (CME) acting on these axial charges induces a net electric quadrupole moment ({\it right}).}
\end{figure}

To estimate the contribution of this mechanism to $\Delta v_2$, we start from the amount of axial charge density created by the non-zero $\vec E\cdot\vec B$ during the lifetime $\tau_E$ of the electromagnetic field,
\begin{equation}
 J^0_A\approx {N_c e^2 \over 2\pi^2}\left(\sum_F q_F^2\right)
EB\tau_E
\approx0.1\, (eE)(eB)\tau_E\,.
\end{equation}
This corresponds to an axial chemical potential of order
$\mu_A\approx j^0_A/\chi_A$ with the axial charge susceptibility
$\chi_A$ whose magnitude we can estimate assuming it is similar to the
vector charge susceptibility $\chi$ known from lattice QCD \cite{Hegde:2008rm}
to be about $\chi\approx  1.0\,T^2$. The CME in Eq.~(\ref{CME}) then produces a
current
\begin{equation}
J={N_c\over 2\pi^2}\left(\sum_F q_F^2\right)\mu_A (eB)\approx 10^{-2}\,T^{-2}\,(eE)(eB)^2\tau_E\,,
\end{equation}
which would result in charge separation $\Delta Q\approx J A\tau_J$
with $A\approx R\tau_E$ as before. We could also expect that the
duration of the chiral magnetic current $\tau_J$ is approximately given by the
duration of the magnetic field $\tau_J\approx\tau_B$.  
Putting this together and
estimating the electric and magnetic fields again as
 $eE\approx eB\approx  \gamma{Z \alpha/  R^2}$  one finds for $\Delta
 v_2\approx\Delta Q/N_{\rm tot}$
\begin{equation}
\Delta v_2
\approx  10^{-2}\, \frac{ (eE)(eB)^2\tau_E^2\tau_B R}{N_{\rm tot}T^2}
\approx 10^{-2}\, \frac{(Z\alpha)^3\tau_E^2\tau_B\gamma^3}{N_{\rm tot}R^5 T^2}
 \approx10^{-7} \left(400\, {\rm MeV}\over T\right)^2 
 \left(10^3\over N_{\rm tot}\right)\,,\label{result22}
\end{equation}
where we used
$\tau_E\approx\tau_B\approx {2R/\gamma}$ and $R\approx 7$ fm as
before. We see that the magnetically induced electric
quadrupole effect is negligible at RHIC energies compared to the direct
electric quadrupole effect in Eq.~(\ref{eq:dv2-2}). However, the
direct effect in Eq.~(\ref{eq:dv2-2})
decreases with energy $\sqrt s$, while the magnetically induced effect
in Eq.~(\ref{result22}) is energy
independent. 

In addition, the magnetically induced electric
quadrupole is sensitive to the lifetime of the  magnetic field $\tau_B$.
For example, if  this time turns out to be much greater than our
estimate $\tau_B=2R/\gamma$ (due to the conductivity of the medium, as
in Ref.~\cite{Tuchin:2013ie}), that effect may be
significantly larger than our estimate in Eq.~(\ref{result22}) and
could possibly compete with the direct electric dipole effect in
Eq.~(\ref{eq:dv2-2}) at sufficiently high $\sqrt s$:
\begin{equation}
\Delta v_2
 \approx10^{-5} 
\left(\tau_B\over 2 {\rm ~fm}\right)
 \left(eB\over 1 {\rm ~fm}^{-2}\right)
\left(400\, {\rm MeV}\over T\right)^2 
 \left(10^3\over N_{\rm tot}\right)\,.\label{result33}
\end{equation}
Comparing this with the estimate of the direct
electric quadrupole effect (\ref{eq:dv2-2}) we see that, while at top
RHIC energy the direct electric quadrupole dominates, the two effects
could possibly become comparable in magnitude ($10^{-5}$) at LHC energies due to
different $\sqrt s$ dependence.

\section{Discrete symmetries}

We end this note by discussing two discrete symmetries, the charge
conjugation $C$ and the $180^\circ$  rotation (or reflection) in the transverse plane,
$R_\perp$, useful for classifying
 possible mechanisms contributing to the observables: the slope $r$ and the intercept $\ctwo$.
Under $C$, 
\begin{equation}
  \Delta v_2(0)\to-\Delta v_2(0)\quad,\quad A_\pm\to-A_\pm\,,
  \ee
therefore $r$ is $C$-even whereas $\ctwo$ is $C$-odd. 

Since QCD is $C$-invariant and the source of $C$-violation is in the
initial condition, one concludes that the physics of $\ctwo$
must have its origin in the initial charge asymmetry, such as the electromagnetic charge of the
heavy-ions (or the isospin asymmetry as in Ref.~\cite{Dunlop:2011cf}). On the other hand, the slope $r$ could possibly receive contributions
from other effects unrelated to the charge asymmetry of the initial conditions
(as, e.g., in Ref.~\cite{Bzdak:2013yla}).

Under the $R_\perp$ rotation (reflection) in the transverse plane, both $r$ and $\ctwo$ are even, since $v_2$ is $R_\perp$-even. The pattern of the electric field is $R_\perp$-even, while the magnetic field pattern
is $R_\perp$-odd. 
In conjunction with the $C$-parity discussion above, this tells
us that the intercept $\ctwo$ can be linearly proportional to the electric field, but not the magnetic field.
Indeed, our first source can be viewed as being linear in the electric
field $E$, whereas the second source should be considered as an
$(EB)B\sim E B^2$ effect.

\section{Summary and Conclusion}

We identified two possible sources of the charged
elliptic flow $\Delta v_2(A_\pm)$ at zero charge asymmetry, i.e.,
$\ctwo$. The most straightforward source is the quadrupole pattern of
the electric fields created by the relativistic heavy ions in the central
overlap region. Although these fields last a short time, their
magnitude is extremely large due to the well-known Lorentz
contraction. Our order of magnitude estimates suggest that the
contribution of this effect at top RHIC energy is comparable with the
observed value of $\ctwo$ and that it should be inversely proportional
to the collision energy $\sqrt s$, dropping by an order of magnitude
at LHC energies. 

The simultaneous presence of electric and magnetic
fields in the overlap region  via the chiral magnetic effect
provides another source of the quadrupole
moment at zero charge asymmetry. The magnitude of this effect is less
sensitive to the collision energy. It depends, however, on the lifetime of
the magnetic fields, which could be longer than that of the electric
fields. We estimate that this effect is negligible at RHIC energies
but may become comparable with the direct electric quadrupole at LHC.

We stress that the two mechanisms we point out in this paper give rise
to the same (positive out of plane) sign of the electric quadrupole moment, which
makes the prediction for the sign of the effect robust.
However, a reliable quantitative prediction of the magnitude would
clearly require a realistic simulation including spatial variations
and fluctuations of the
electromagnetic fields
\cite{Bzdak:2011yy,Deng:2012pc,Bloczynski:2012en}.

\section*{Acknowledgments}

We thank Adam Bzdak, Olga Evdokimov, Jinfeng Liao, Todd Springer, and
Yi Yin for helpful discussions. The work of M.S.
is supported by the DOE Grant No. DE-FG02-01ER41195.

 \vfil

\end{document}